# A win-win monetary policy in Canada

Oleg Kitov, Department of Economics, University of Oxford

Ivan Kitov, Institute for the Geospheres' Dynamics, RAS

Abstract
The Lucas critique has exposed the problem of the trade-off between changes in monetary policy and structural breaks in economic time series. The search for and characterisation of such breaks has been a major econometric task ever since. We have developed an integral technique similar to CUSUM using an empirical model quantitatively linking the rate of inflation and unemployment to the change in the level of labour force in Canada. Inherently, our model belongs to the class of Phillips curve models, and the link between the involved variables is a linear one with all coefficients of individual and generalized models obtained by empirical calibration. To achieve the best LSQ fit between measured and predicted time series cumulative curves are used as a simplified version of the 1-D boundary elements (integral) method. The distance between the cumulative curves (in $L^2$ metrics) is very sensitive to structural breaks since it accumulates true differences and suppresses uncorrelated noise and systematic errors. Our previous model of inflation and unemployment in Canada is enhanced by the introduction of structural breaks and is validated by new data in the past and future. The most exiting finding is that the introduction of inflation targeting as a new monetary policy in 1991 resulted in a structural break manifested in a lowered rate of price inflation accompanied by a substantial fall in the rate of unemployment. Therefore, the new monetary policy in Canada is a win-win one.

## Introduction

The Bank of Canada was one of the first worldwide to announce the policy of inflation targeting between 1 and 3 percentage points per year (Bank of Canada 2010). This objective was originally articulated in 1988 and this new monetary policy was formally introduced in 1991. The goal to retain the level of price inflation in the designated range may introduce a measurable disturbance in a given economy and affect the links between major macroeconomic variables, as was explicitly indicated by Lucas (1976). Then, the basis of price inflation targeting might be corrupted since inflation sensitivity to other macroeconomic variables may change. In this case, one would have observed "structural breaks" in the relationships between such macroeconomic time series as inflation and unemployment (or some other measures of economic activity), i.e. sudden changes in the related empirically derived coefficients (Hostland 1995).

In this paper, we quantitatively estimate and statistically characterize the evolution of several deterministic relationships between inflation and unemployment in Canada. One of our major objectives is to reveal and better estimate in time and amplitude the structural break potentially associated with the introduction of inflation targeting. For this purpose, we have adapted from physics and engineering the method of boundary elements in its simplified form of cumulative curves, which is complementary to the econometric techniques based on dynamic and differentiated time series. In some cases, the dynamic approach is subject to type I spurious regression damaging the estimation of actual links between time series (Chiarella and Gao 2002).

Having estimated with the 1D boundary elements method a number of statistically reliable deterministic models of inflation and unemployment (Kitov and Kitov 2010) we are able to find the trade-off between these variables, which may be best expressed in cumulative values (integrals). Page (1954) introduced a technique for univariate time series, which comprises the statistical basis for our empirical approach. This is a well-known CUSUM (cumulative sum) control chart. We use essentially bivariate and trivariate deterministic model with nonstationary time series and have to calibrate the model together with testing for structural breaks. Therefore, it is instructive to plot both (measured and predicted) times series instead of their demeaned difference. However, all statistical inferences related to the CUSUM method can be applied one-to-one to the model residual.

Any structural break is (by definition) accompanied by the change in relevant model coefficients. When both sides of a given bivariate relationship are integrated over time the structural break must manifest itself in the divergence of integrals starting from the break point. Statistically, this approach is potentially a more reliable one than those based on specific features of dynamic time series. Firstly, all uncorrelated noise is suppressed by destructive interference. Secondly, any amplitude dependent systematic error is compensated by a proportional shift in the slope and all amplitude independent systematic errors add up to the free term. It is worth noting that nobody is able to measure true values of such macroeconomic variables as inflation and unemployment; they both depend on definitions which are never perfect. Hence, one always has systematic errors in these time series, which are not easy to handle in the dynamic representation since they may introduce an amplitude-dependent bias. Thirdly, the effect of the change in coefficients is steadily amplified (accumulated) by constructive interference with increasing signal to noise ratio. This effect is crucial for statistical inferences. Literally, the integral curves representing both sides of the equation diverge at a constant rate after the year of structural break.

For our model, all three benefits are working together. An additional (but not uncommon for econometrics) benefit consists in the fact that the cumulative sums of price inflation and the change rate of labour force are represented by actually measured overall price and labour force. Since the accuracy of price and labour force measurements is relatively high and approximately time independent the cumulative error terms in the time series of inflation and the change rate of labour force must always add up to the level of the measurement accuracy, i.e. asymptotically converge to a zero mean. This is a strong constraint on the error term in standard statistical inferences. In reality, all past values of labour force and prices are routinely revised with every new measurement to match the newly measured value.

However, all these benefits are conditional on the presence of a deterministic link. When a purely statistical link between two stochastic variables is integrated, the uncorrelated error term creates a stochastic trend, the systematic error correlates with the stochastic trend and cannot be separated from it, and the divergence between the integral curves after the break is not a quasi-linear one. This is the reason why econometricians do not use CUSUM. There is a slight prejudice against deterministic links in econometrics.

Fortunately, a variety of actual measurements reported by developed countries allow distinguishing between deterministic and stochastic links (Kitov 2007a; Kitov and Kitov 2010). It was found and proved by strict and extensive statistical and econometric tests (Kitov, Kitov, and Dolinskaya 2007) that the evolution of price inflation and the rate of unemployment is driven by the change in the level of labour force. For Canada, we estimated these links several years ago (Kitov 2007b), without structural breaks however. Thus, we can validate the previous models using new data and refine them allowing for structural breaks.

Several years ago we introduced a concept linking by linear and lagged relationships price inflation and unemployment in developed countries to the change rate of labour force (Kitov, 2006). Corresponding model is a completely deterministic one with the change in labour force being the only driving force causing all variations in the pair unemployment/inflation. Since 2006, many empirically estimated models have been tested econometrically (conditional on the length of time series) and the presence of cointegrating relations has been demonstrated. Formally, our model is a somewhat degenerate version and a marginal extension of such economic/econometric models as the conventional Phillips curve or the new Keynesian Phillips curve. For example, among the diversity of economic/financial variables used by Stock and Watson (2003, 2008) as predictors of inflation, the set included approximately 200 time series, the change in labour force was absent. Thus, it was instructive to extend this set with labour force and to conduct a similar statistical investigation.

We have revealed for many developed countries (the USA, Japan, France and Germany among others) that, in purely econometric terms, this rather countable than measurable macroeconomic predictor is characterized by an extraordinary (relative to other tested parameters) power and inflation is not "hard to forecast", as concluded by Stock and Watson (2007). The change in labour force in the biggest developed economies is so good a predictor that there is no need to use autoregressive properties of inflation and other variables. In this sense, our model is fully deterministic and the model residuals are related to measurement errors not to stochastic properties of the involved processes.

Here we have to notice that such an extensive usage of autocorrelation in the modelled time series implies that the researcher does not expect any other macroeconomic variables to be involved. Mathematically, it is a flawed way of quantitative analysis – autocorrelation terms severely mask any real driving force since the modelled time series is decomposed into a set of non-orthogonal functions (variables). The long history of econometric research has unambiguously demonstrated that the AR and similar statistical models are able to describe observations only superficially and suppress the signals from actual sources of inflation. Even with wrong predictors, the Phillips curve approach works best when inflation changes very fast and autocorrelation is highly deteriorated, as was observed between 1974 and 1994 in the U.S. (Stock and Watson 2008). Then real forces reveal themselves: the change in labour force explains 80 to 90 per cent of the variability in the rate of inflation, with no AR terms.

The remainder of the paper is organized in four sections. Section 1 formally introduces the model as obtained and statistically tested in previous studies and highlights its major features different from those in the extensive literature on inflation and unemployment. This Section also presents and statistically characterizes various estimates of inflation, unemployment and labour force in Canada.

In Section 2, the linear link between labour force and unemployment is modelled using annual measurements of both variables. Instead of poorly constrained linear regression methods we apply a simplified version of the 1-D boundary element method – cumulative curves with the *LSQ* minimization. The integral approach allows distinguishing a structural break near 1990 when the predicted and observed curves start to diverge. In order to retain the convergence between the curves intact after 1990, the model coefficients are changed to minimize the *LSQ* residual. Section 3 is devoted to the link between the rate of inflation and labour force. We also use the method of cumulative curves in order to estimate all empirical coefficients and the year of structural break.

Finally, Section 4 presents the generalized link between inflation, unemployment and labour force which is characterized by the absence of any structural breaks. The best fit model provides an accurate prediction of inflation as a function of labour force and unemployment without changing coefficients.

1. **The model and data**

In its original form, the model was revealed and formulated for the United States (Kitov 2006). The root-mean-square forecast error (RMSFE) of inflation at a 2.5 year horizon was of 0.8% between 1965 and 2004. Thus, our model outperforms by a large margin any other economic and/or financial model of inflation; at least those presented by Stock and Watson (2008). Well-known non-stationary behaviour of all involved variables required testing for the presence of cointegrating relations (Kitov, Kitov, and Dolinskaya 2007). Both, the Engle-Granger and Johansen approaches have shown the existence of cointegration between unemployment, inflation and the change in labour force, i.e. the presence of long-term equilibrium (in other words, deterministic or causal) relations. Because the change in labour force is likely a process with a strong stochastic component and it drives the other two variables (with significant lags) they also can exhibit formal features of the underlying stochastic process, at the same time being fully deterministic ones.

Here, we generally follow the original concept introduced by A.W. Phillips (1958) but suppose that price inflation and the rate of unemployment in a given developed country have to share the same driving force, and thus, there is a trade-off between them. Mathematically, inflation and unemployment are both linear and potentially lagged functions of the change rate of labour force:

$$\pi_t = a_1 l_{t-i} + a_2 \quad (1)$$
$$u_t = b_1 l_{t-j} + b_2 \quad (2)$$

where $\pi_t$ is the rate of price inflation at (discrete) time $t$, as represented by some standard measure such as the GDP deflator (DGDP) or consumer price index (CPI); $u_t$ is the rate of unemployment at time $t$, which also may have varying definition and measuring procedures; $l_t = dlnLF(t)/dt$ is the log approximation to the growth rate of the level of labour force at time $t$, $LF(t)$; $i$ and $j$ are the (not negative) time lags; $a_1$, $b_1$, $a_2$, and $b_2$ are country specific coefficients, which have to be determined empirically in calibration procedures. There is no error term in (1) and (2) since the left- and right-hand sides must converge for a deterministic relationship by definition, with the error term fully related to measurement errors and its cumulative sum having a zero mean asymptote. Free term $a_2$ in (1) might replace the notion of "intrinsic inflation persistence" (Benati, 2009). The rate of inflation with zero driving force ($l_t=0$) is constant, but this rate is not necessary a policy independent one. The same statement is valid for $b_2$ - it is the rate of unemployment in the absence of any change in labour force.

All coefficients in (1) and (2) are subject to variations through time for a given country. The major source of such variations is numerous revisions to definitions and measurement methodologies of the studied variables, i.e. variations in measurement units. For example, the change of employment definition from $n$ hours per week to $m$ hours, where $n>>m$, must induce a tangible change in the number of employed, and thus, in the rate of unemployment. Since only units of measurements are changed, i.e. the portion of the true value, the corresponding shift in coefficients in (1) and (2) is an artificial structural break, like mile to kilometre conversion. The introduction of monetary policy aimed at strong suppression of money supply, as implemented by the French central bank, is also able to change all coefficients (Kitov 2007). This is an example of an actual structural break associated with monetary policy, in sense of Lucas. We are looking for actual structural breaks in Canada, and thus, have to be very careful with data incompatibility over time.

Linear relationships (1) and (2) define inflation and unemployment separately. These variables are two indivisible manifestations or consequences of a unique process, however. The process is the growth in labour force which is accommodated in developed economies (we do not include developing and emergent economies in this analysis because they are likely not self-consistent) through two channels and results in the trade-off between inflation and unemployment, as was empirically revealed by A.W. Phillips. The original Phillips curve concept strictly implies that $i \geq j$, i.e. the change in unemployment drives inflation.

Considering the qualitative assumptions behind the quantitative model (1) and (2) we have revealed two processes which accommodate the endogenous (inflow of 16-year-olds and the age-dependent rate of death and labour force participation) or exogenous (international migration) change in labour force. The first process is the increase in employment and corresponding change in personal income distribution (PID). Since the rate of participation in labour is completely defined by real economic growth (Kitov and Kitov 2008) the increase in employment does not depend on inflation and unemployment. Thus, real economic growth involves new persons who obtain new paid jobs or their equivalents and presumably change their incomes to some higher levels. Interestingly, a higher growth rate of real GDP in the U.S. causes a fall in the rate of participation in labour force two years later. This is a counterintuitive observation.

These newcomers do not change the relative distribution of incomes, however. There is a well-established empirical fact that the PID shape in the U.S. does not change with time in relative terms, i.e. when normalized to total population and total income. Therefore, the increasing number of people at higher income levels, as related to the new paid jobs, must be accompanied by a certain disturbance in the nominal PID. (This process is opposite to that behind the original Phillips curve, where the general trade-off between inflation and unemployment is not strictly constrained). This over-concentration (or "over-pressure") of working population in some income bins above its "neutral" long-term value must be compensated by such an extension in corresponding income scale, which returns the PID to its original density. The related income scale stretch (money supply) is the core monetary process behind price inflation. In other words, the U.S. economy needs exactly the amount of money, extra to that related to real GDP growth, to pull back the PID to its fixed shape. The mechanism responsible for the compensation and the income scale stretching may have some positive relaxation time, which effectively separates in time the source of inflation (i.e. the labour force change) and the reaction - the growth in the overall price level.

The second process involves those persons in the labour force who wanted but failed to obtain a new paid job. These people do not leave the labour force and raise the rate of unemployment. Supposedly, they do not change the PID shape because they do not increase their incomes. Therefore, the total labour force change equals the unemployment change plus the employment change, the latter process expressed through lagged price inflation.

In the case of a "natural" or stationary behaviour of the economic system, which is defined as a stable balance of socio-economic forces in the society, the partition of labour force growth between unemployment and inflation is retained through time and the linear relationships hold separately. There is always a possibility, however, to fix one of these two dependent variables. Central banks are definitely able to influence inflation rate by monetary means, i.e. to force money supply to change relative to its natural demand. To account for this effect one has to use a generalized relationship as represented by the sum of (1) and (2):

$$\pi_t + u_t = a_1 l_{t-i} + b_1 l_{t-j} + a_2 + b_2 \qquad (3)$$

Equation (3) balances the change in labour force to inflation and unemployment; the latter two variables may lag by different times. When $i \neq j$, one cannot relate inflation and unemployment for the same year. Theoretically, equation (3) overcomes the Lucas critique – no monetary policy should be able to change the generalized relation between these three macroeconomic variables. The change in inflation is compensated by a proportional change in unemployment, with some time lag, which also can be negative. In practical terms, the importance of (3) is derived from the increasing number of successful prediction of inflation and unemployment in developed countries (Kitov and Kitov 2010). One can rewrite (3) in a form similar to that of the Phillips curve:

$$\pi_t = c_1 l_{t-i} + c_2 u_{t+j-i} + c_3 \qquad (4)$$

where coefficients $c_1$, $c_2$, and $c_3$ might be better determined empirically despite they can be directly obtained from (3) by simple algebraic transformation. The change in labour force is always the measure of economic activity instead of output gap and marginal labour cost.

When i>j, the rate of unemployment mimics the term associated with rational or not fully rational inflation expectations. In any case, the menu cost, distribution of price setting power, nominal rigidities, sticky prices and information and other components of the New Keynesian Phillips curve might have a simple functional form under our framework. Our model puts a strict constraint on the aggregate value of any parameter used under the NKPC framework. In that sense, we do not see any contradiction between our deterministic model and the variety of NKPC models. This is an issue for further theoretical investigations, however.

The principal source of information is the OECD database (http://www.oecd.org) which provides comprehensive data sets on labour force, unemployment, the GDP deflator, and CPI. We also use select estimates reported by the U.S. Bureau of Labour Statistic (http://www.bls.gov) for corroboration of the data on CPI, unemployment and labour force. As a rule, readings associated with the same variable but obtained from different sources do not coincide. This is due to different approaches and definitions applied by statistical agencies. The discrepancy between various estimates of the same variable is often associated with data incompatibility. When two estimates suddenly diverge or start to coincide it is possible to suggest that one of the agencies has adapted a new definition. The diversity of definitions is accompanied by a large degree of uncertainty related to the methodology of measurements. In many cases, this uncertainty cannot be directly estimated but certainly affects the reliability of empirical relationships.

Since there is no full compatibility in definitions and measurement procedures over time all data provided by all statistical agencies have to be checked for artificial breaks. It is crucial to distinguish between these breaks in measuring units and actual shifts in the relationships between the modelled variables. For Canada, the OECD (2008) reports the following:

> *Series breaks*: In January 1976 the following revisions to the labour force survey were implemented: sample increase, from 35 000 to approximately 56 000 households; update of the sample by redesign; introduction of new methodology and procedures at the level of stratification, sample allocation and formation of sampling units; improvement of data collection techniques, quality control and evaluation procedures. Prior to 1966, the survey covers population aged 14 years and over.

Taking into account that the Statistics Canada has been reporting the labour force related time series since 1976, one can expects artificial breaks in corresponding variables in 1966, 1976, and 1990. There were several potentially influential revisions to the Labour Force Survey prior to 2000 (e.g. 1996 and 2000), as described by the Statistics Canada (2011). One has to bear them in mind when searching for structural breaks. In general, any deterministic model linking the originally reported estimates of labour force to inflation and unemployment would experience significant difficulties which can be resolved only by the appropriate shift in empirical coefficients in the break years. One cannot exclude other influential revisions after 2000.

Figure 1 displays the evolution of two principal measures of price inflation – the GDP deflator, *DGDP*, and consumer price index, *CPI*. Both variables were published by the OECD. We presume the DGDP as a better representative of price inflation in a given country since it includes all prices related to the economy. The overall consumer price index is fully included in the DGDP and its behaviour is usually more volatile as representing only a (larger part) of the economy. Since labour force and unemployment do characterize the entire economy it is methodically correct to use the price deflator for quantitative modelling. Figure 1 shows that the difference between the CPI and DGDP estimates can reach several percentage points and their major peaks are not proportional in amplitude, e.g. 1974, 1982 and 1992. However, there are periods of coherency.

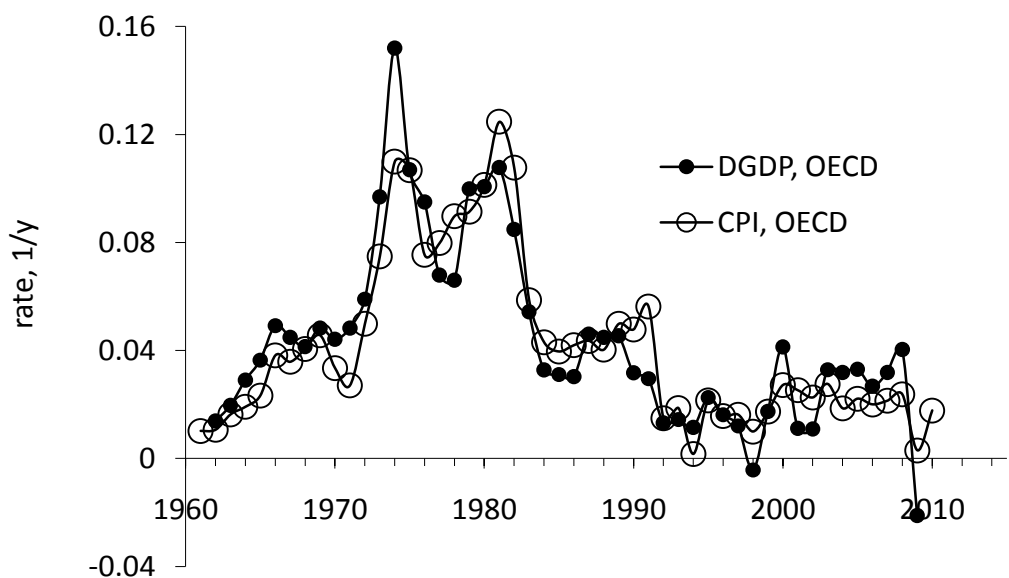

Figure 1. Price inflation: comparison of CPI and the GDP deflator in Canada, both reported by the OECD.

For the period between 1962 and 2009 (48 readings) the mean rate of CPI inflation is $0.043y^{-1}$ (±0.031) and $0.044y^{-1}$ (±0.033) for the GDP deflator. Thus, the DGDP in Canada is characterized by a higher volatility as associated with the peak in 1974. The rate of inflation fell to the level of $0.04y^{-1}$ in the beginning of 1980s and then to $0.02y^{-1}$ in 1991, i.e. after the introduction of inflation targeting. It has been oscillating around this level since. Between 1974 and 1983, inflation was almost everywhere above 6% per year.

Any macroeconomic variable is subject to measurement uncertainty and bias. Rossiter (2005) has considered the bias in the Canadian CPI by examining four different potential sources: commodity substitution bias, outlet substitution bias, quality change bias, and new goods bias. He found that the total measurement bias has increased only slightly in recent years to 0.6 percentage points per year ($0.006y^{-1}$), and is low when compared with other countries.

In order to establish a reliable link with the labour force, one needs to estimate basic statistical properties of relevant time series. The order of integration can be determined using unit root tests applied to the original series and their progressive differences. As a rule, the rate of inflation in developed countries is an $I(1)$ variable. Canada is not an exception, as Table 1 clearly demonstrates. In particular, we report the results of the following tests: the augmented Dickey-Fuller (ADF), the DF-GLS, and the Phillips-Perron (PP) test. The *DGDP* series consists of 48 readings (between 1962 and 2009) and the CPI series has 50 readings, both definitely have a unit root. The first differences (*dDGDP* and *dCPI*) are characterized by the absence of unit roots, and thus, the original time series is integrated of order 1. For the period between 1963 and 2009, the naive predictions of inflation at a one-year horizon have the standard deviations of $0.020y^{-1}$ and $0.016y^{-1}$, respectively.

Table 1. Results of unit root tests for the original time series and their first differences.

|  | ADF | DF-GLS lag=2 | PP $z(\rho)$ | $z(t)$ |
|---|---|---|---|---|
| CPI | -1.63 (-3.59) | -1.56 (-3.77) | -6.13 (-18.76) | -1.72 (-3.59) |
| dCPI | -5.61* (-3.61) | -4.79* (..) | -34.84* (-18.63) | -5.51* (-3.61) |
| DGDP | -1.76 (-3.60) | -1.76 (..) | -7.69 (-18.70) | -1.80 (-3.60) |
| dDGDP | -5.49* (-3.61) | -3.77 (..) | -36.62* (-18.63) | -5.26* (-3.61) |
| UE | -1.59 (-3.59) | -1.55 (..) | -6.18 (-18.76) | -1.77 (-3.59) |
| dUE | -5.07* (-3.61) | -3.97* (..) | -32.91* (-18.63) | -4.94* (-3.61) |
| dLF/LF | -2.71 (-3.59) | -1.87 (..) | -12.76 (-18.76) | -2.64 (-3.59) |
| d(dLF/LF) | -9.32* (-3.59) | -4.37*(..) | -58.41* (-18.76) | -9.69* (-3.59) |

Figure 2 depicts two estimates of the rate of unemployment as reported by the OECD and the U.S. BLS. Surprisingly, the difference between these curves is almost negligible, but clearly demonstrates two artificial breaks in 1966 and 1976. There are two sharp peaks – in 1984 and 1994. The highest rate of unemployment in Canada was at the level of 11.4% (OECD definition) in 1984, and the lowermost one was around 3.8% in 1967. Table 1 shows that the $u_t$ series is likely integrated of order 1.

The rate of change in labour force, $l_t$, in Figure 3 also has two representations: the OECD and BLS. Both time series are in an excellent agreement and the rate of change is practically identical, except in 1967 and 1976. These are the years of revision to labour force related definitions, with the spikes potentially damaging all *LSQ*-based statistical inferences. Table 1 indicates that the annual estimates of labour force represent rather an *I*(1) process. The augmented Dickey-Fuller and Phillips-Perron both reject the null hypothesis of a unit root. However, the DF-GLS test does not reject the null for all lags above 1 (only lag 2 is shown in the Table). The first difference, $dl_t$, is an *I*(0) process with all tests rejecting the null hypothesis of a unit root.

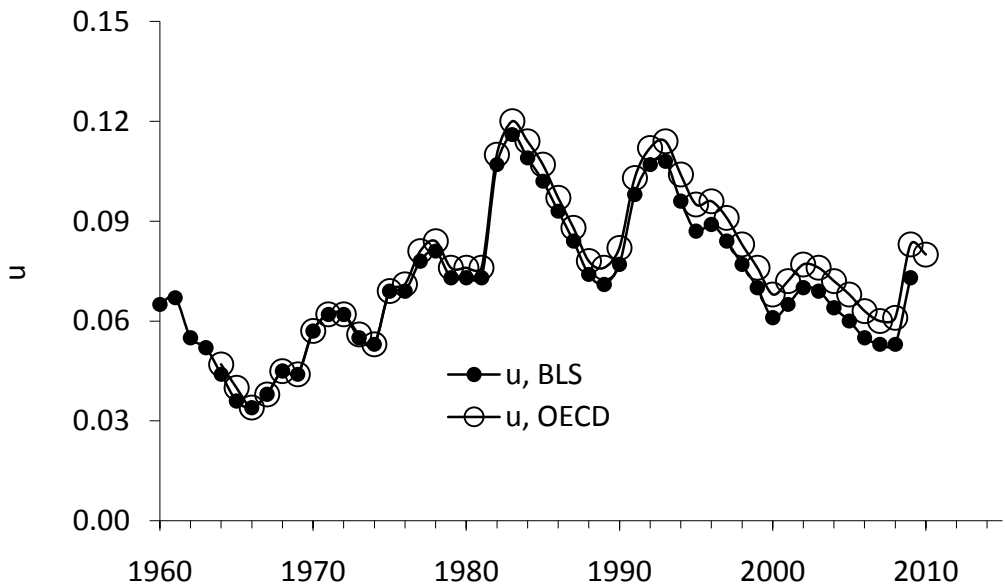

Figure 2. Comparison of two estimates of unemployment according to the U.S. BLS and OECD definitions.

Since linear functional dependences between the three involved *I*(1) processes are estimated later on, econometric analysis requires specific tests for cointegration. We would not like the results of our statistical estimates to be biased by stochastic trends, as was originally found by Granger (1981) for various economic series.

The main task of this study is to accurately characterize in time and size the structural break associated with the new monetary policy formally introduced in 1991. For this purpose, we use the best fit empirical relationships between inflation/unemployment and labour force. By definition, all relationships are linear and potentially lagged. Technically, this task does not seem to be a difficult one since we use cumulative curves, accompanied by the residual minimization in the $L^2$-metric, instead of linear regression applied to the dynamic series. (The latter method provides heavily biased estimates of the slope when both variables have high uncertainties.) We have also tried using the $L^1$-metrics as applied to the annual and cumulative curves. However, the overall performance of the $L^2$-metrics makes it a more appropriate for the cumulative curves where large-amplitude outliers are almost absent.

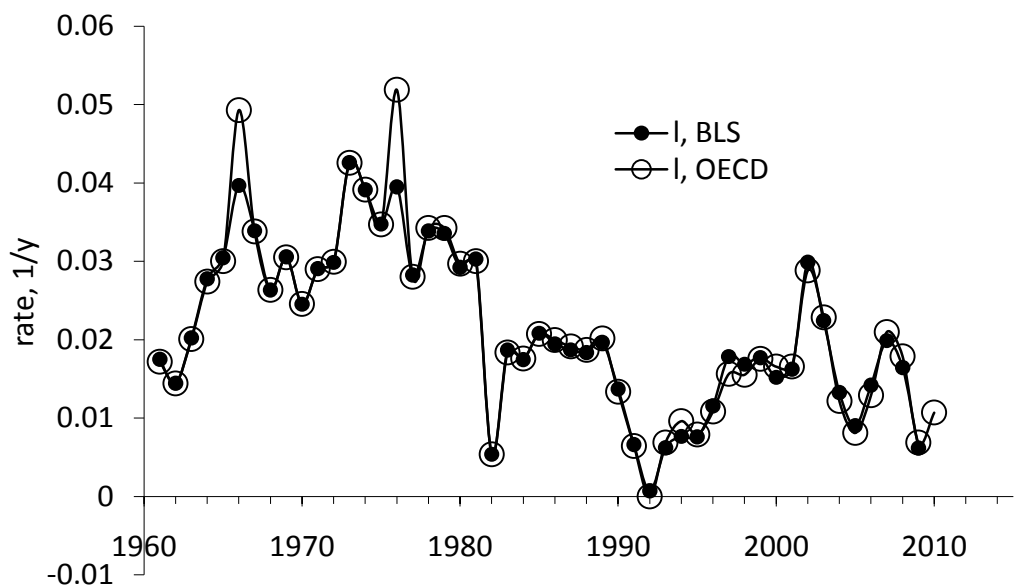

Figure 3. Comparison of two estimates of the change rate of labour force level – as reported by the OECD and BLS.

## 2. Unemployment as a linear function of the change in labour force

In our previous paper on inflation/unemployment in Canada we demonstrated the absence of the Phillips curve in its original form (Kitov 2007b). From the absent Phillips curve it is only one little step to the dependence of unemployment on the change in labour force: we are looking for a driving force. Actually, we replace the rate of inflation in the Phillips curve with the rate of labour force change. Then, we have to estimate both coefficients in (2). The estimation method is enhanced relative to our previous studies – the best overall fit is sought by the least squares method as applied to the cumulative curves. In addition to the formal *LSQ* minimization of the model error we have introduced a freely varying break year in the model. The break should be within 4 years around the assumed year (1991). By definition, the final break year has to provide the lowermost RMS residual. All in all, the best- fit equations and the break year are as follows:

$$u_t = -2.574 l_t + 0.155; \ t<1990$$
$$u_t = -2.852 l_t + 0.122; \ t\geq 1990 \quad (5)$$

with the break in 1990. The change in the break year is likely explained by the influence of measurement noise. As an alternative, one may guess that the Bank of Canada introduced inflation targeting in some testing regime a year before the official start. Figure 4 illustrates the behaviour of the dynamic and cumulative curves. We have intentionally extended the model for the period before 1990 into the 1990s and 2000s in order to demonstrate the structural break in 1990. The cumulative curves clearly deviate since 1990.

In the previous paper, coefficients in (5) were as follows: $b_1$=-2.1 and $b_2$=-0.12. The absolute value of the slope was underestimated since the original equation had to fit the whole period between 1969 and 2004. The visual-fit approach and the absence of the structural break made this task a difficult one. The intercept was estimated with a higher accuracy, at least after 1990. Overall, the original model gave a resonable first approximation.

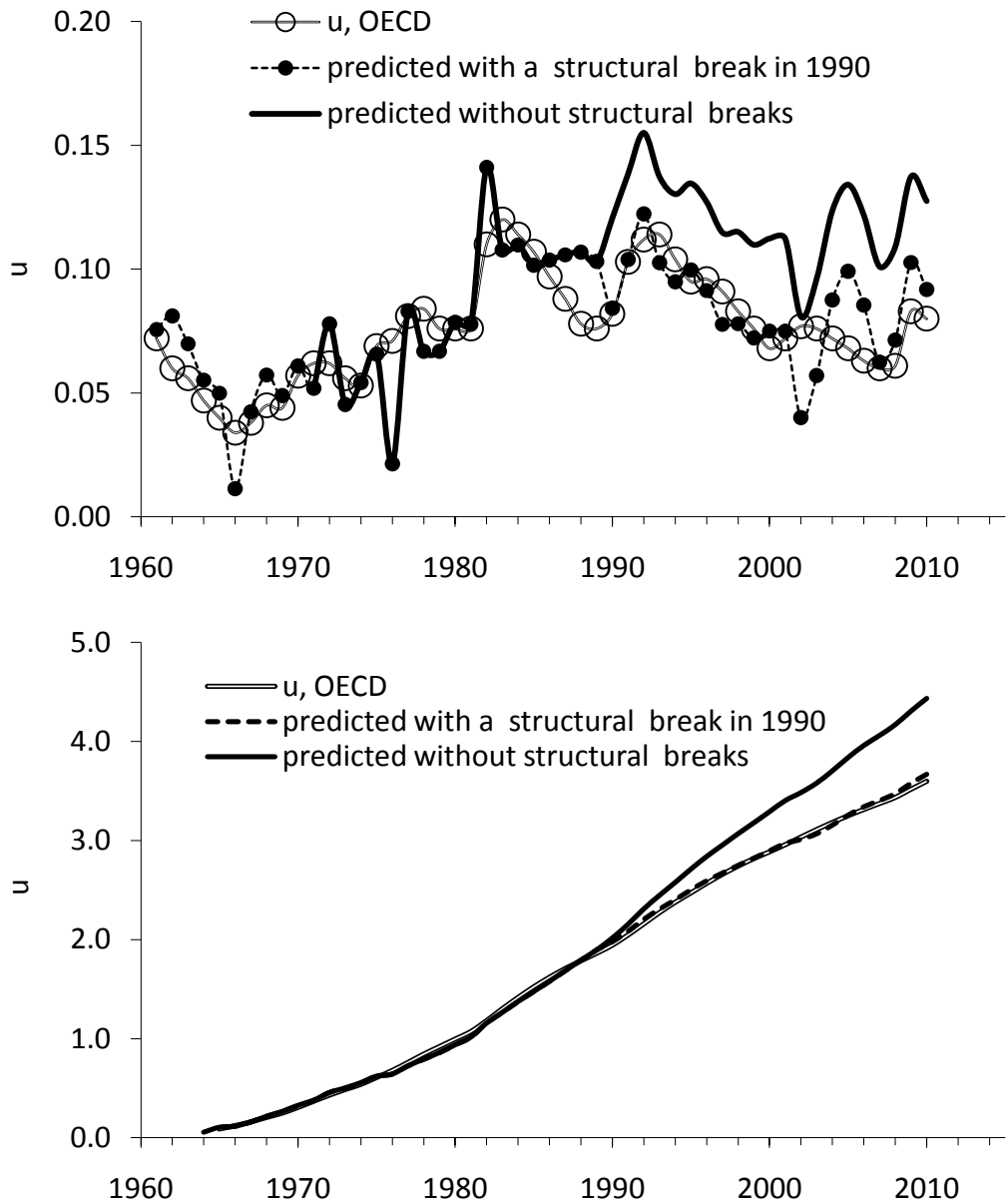

Figure 4. Annual estimates of the rate of unemployment in Canada: measured vs. predicted from the change in labour force.

The slope in (5) is always negative. Therefore, any increase in the level of labour force is reflected in a proportional and simultaneous fall in the rate of unemployment. This is a fortunate link – more people in work force is equivalent to less unemployed. However, when the level of labour force does not change with time the rate of unemployment is very high – around 12% (after 1990). Canada has to keep a higher rate of labour force growth in order to retain the rate of unemplyment at low level.

All in all, the model obtained for the previous period diverge after 1990. The difference between the cumulative curves is positive and increases with time. This is a favourable outcome of the new monetary policy – it has reduced the rate of unemployment by approximately 3.6% relative to that predicted by the relationship valid before 1990 (see Figure 5). For the economic theory, it is likely an unexpected result since the Phillips curve implies an opposite behaviour.

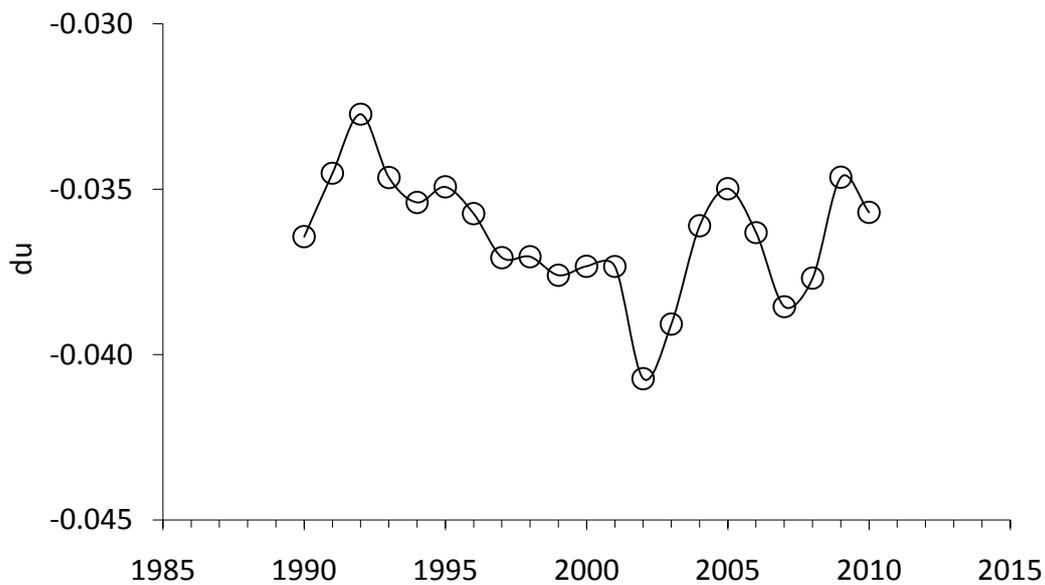

Figure 5. The difference between the observed rate of unemployment in Canada and that predicted by the relationship valid before 1990, i.e. the expected rate of unemployment.

Relationship (5) implies a nonlinear dependence on the rate of particiaption in labour force. For a given absolute change in the level of labour force, the reaction of unemployment will be different for the rate of participation 56% (1964) and 67.7% (2008). The higher is the participation rate the lower is the change rate, $l_t$, and thus, the change in the rate of unemployment. Actually, the participation rate in Canada has been increasing since 1995 to a very high level of 67.7%. It will be a difficult task to retain the rate of unemployment at the current level – it is likely that the rate of participation is approaching the peak level and will start to decline in the near future.

**3. Inflation as a linear function of the change in labour force**

The existence of a deterministic link between labour force and price inflation has also been proved for many countries. We are following the same estimation procedure as for unemployment above, i.e. the method of cumulative curves enhanced by the *LSQ* minimization of the model error and the break year freely varying around 1991.

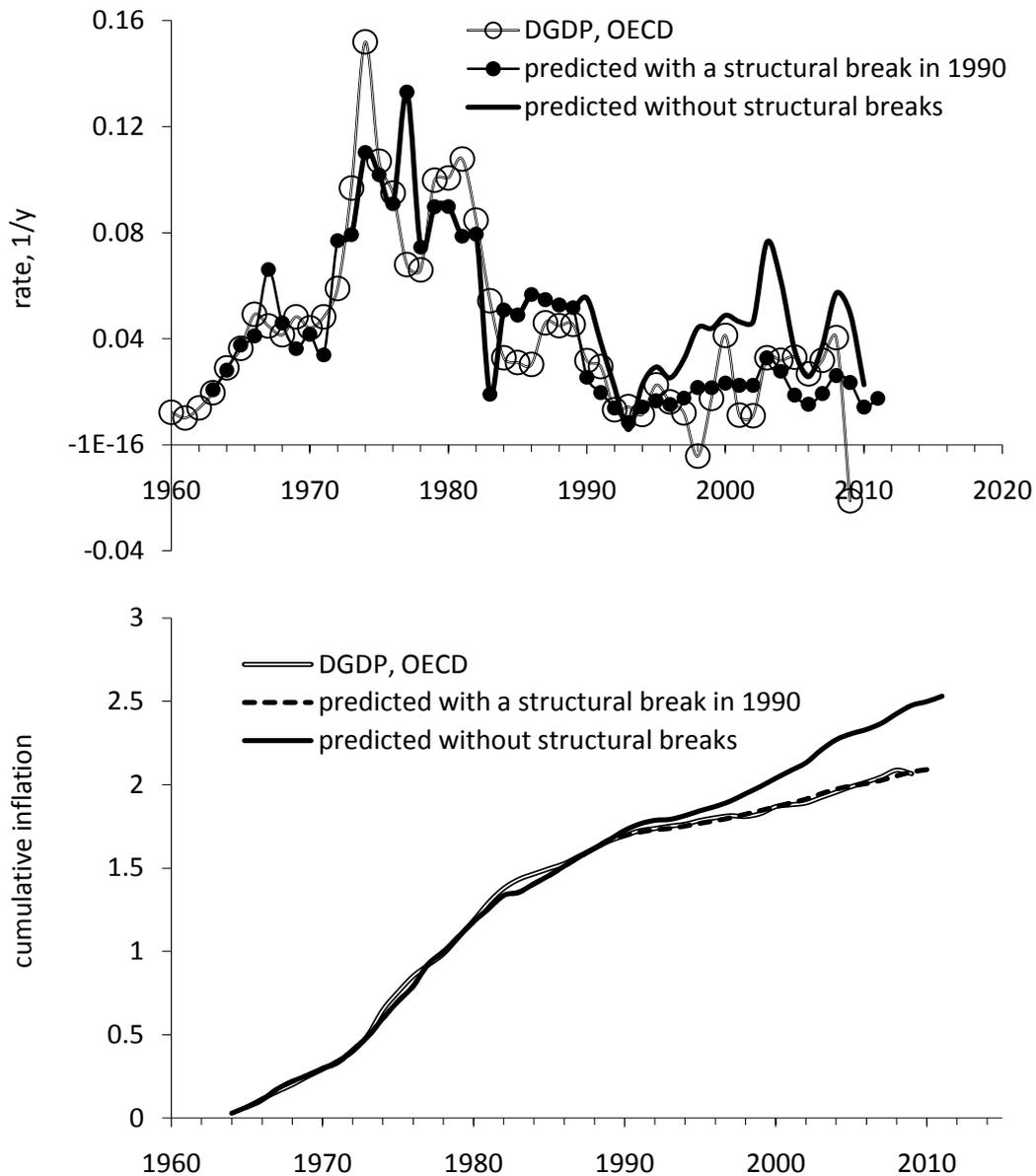

Figure 6. Modelling the annual and cumulative GDP deflator as a function of the change rate of labour force.

We start with annual readings of the GDP deflator reported by the OECD. The minimization procedure with the start break year in 1991 (the change in monetary policy) and zero lag (varied between zero and 5 years) has finally determined the data-driven break year in 1990 and the lag of 1 year:

$$DGDP_t = 2.453 l_{t-1} + 0.0052;\ t<1990$$
$$DGDP_t = 0.842 l_{t-1} - 0.0085;\ t\geq1990 \qquad (6)$$

Figure 6 displays the observed DGDP curve and that predicted according to (6). The cumulative curves are in a good agreement. For the period between 1964 and 2009, the coefficient of determination is very high: $R^2_{dyn}$=0.70 and $R^2_{cum}$= 0.999, respectively. Relationship (6) does not use any past or future values of inflation (i.e. there are no AR terms) which usually bring between 80% and 90% of the explanatory power in the mainstream models (Piger and Rasche 2006; Rudd and Whelan 2005; Stock and Watson 2008). Hence, this link is almost a deterministic one. The cumulative curves demonstrate that the left- and right-hand sides in (6) converge with time.

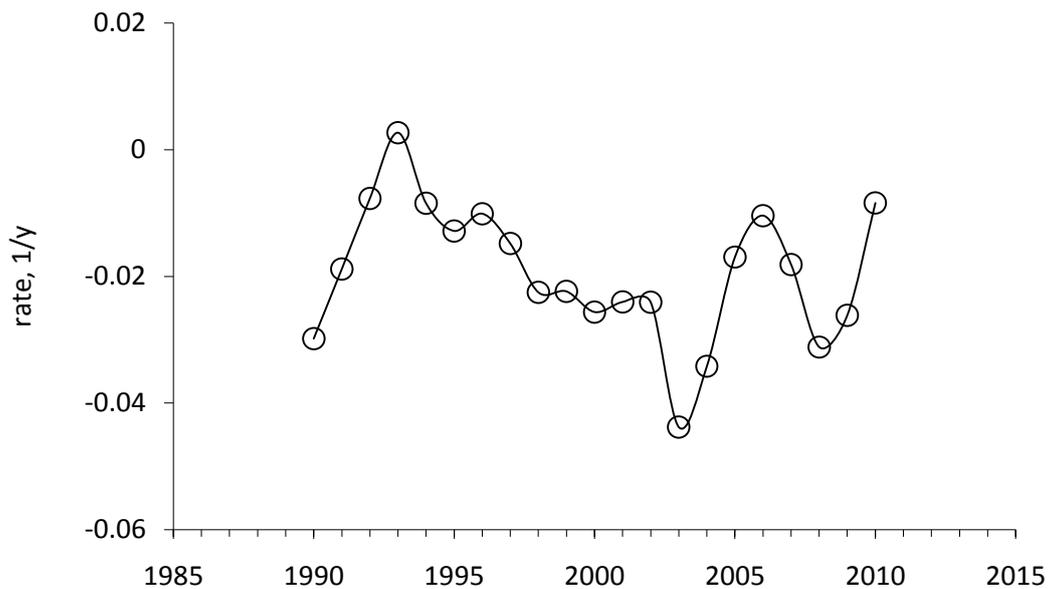

Figure 7. The difference between the measured and expected GDP deflators. The mean value between 1990 and 2009 is -0.019y$^{-1}$.

Figure 7 depicts another benefit of the new monetary policy - the rate of price inflation has been reduced by 1.9% per year on average since 1991 relative to that predicted by the early model. Hence, the policy has a significant effect on the price growth in Canada and the Lucas critique was well justified by the reduced rate of unemployment. Unfortunately, this win-win policy is not the only possible outcome of inflation targeting. In France, a similar (with strict constraints on the level of money supply, however) monetary policy introduced in 1994 has resulted in an opposite reaction of unemployment (Kitov 2007a). On average, the observed rate has been approximately 5% above that predicted by the model valid before 1995. This was not a win-win game.

Figure 8 illustrates the expected benefits of the cumulative approach. The absolute and relative errors decrease with time. Despite the annual levels of price and labour force are not measured more accurately with time the relative change in the level is measured with an increasing accuracy due to the increasing denominator. As a consequence, the observed and predicted cumulative curves, i.e. the overall changes in price and labour force, do converge. They become indistinguishable. Taking into account the lead of the change in labour force by 1 year and its independence on the future rate of inflation one can suggest that there exists a deterministic link between them.

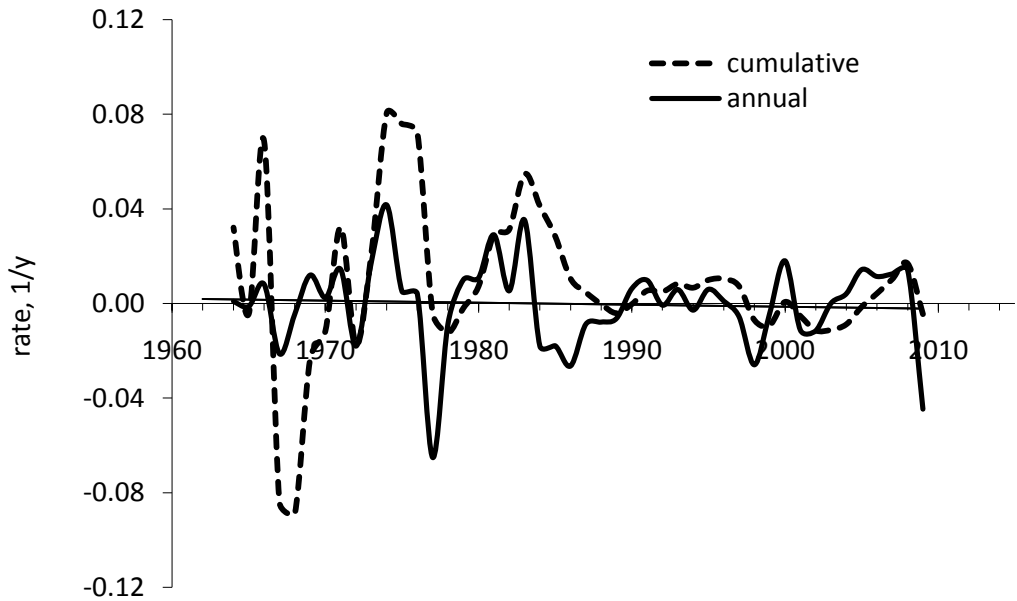

Figure 8. Absolute dynamic and relative cumulative modelling error for the inflation in Figure 6.

The annual time series are relatively short with only 48 readings between 1962 and 2009. Small samples do not guarantee higher confidence of statistical results. However, we have carried our formal tests for cointegration. First, we have tested the differences between the dynamic and cumulative curves, i.e. the model residuals. For the annual residuals, the augmented Dickey-Fuller test is -5.36 with the 1% critical value of -3.61, i.e. the null of a unit root is rejected. The Phillips-Perron tests gives $z(\rho)$=-35.33* (-18.56) and $z(t)$=-5.23* (-3.61). The DF-GLS test rejects the null for all lags between 1 and 5, except 4.

For the residuals of the cumulative model, the augmented Dickey-Fuller test is -4.16* with the 1% critical value of -3.61. The Phillips-Perron tests gives $z(\rho)$=-26.05* (-18.56) and $z(t)$=-4.18* (-3.61). The DF-GLS test rejects the null for all lags between 1 and 5, except 4. Therefore, both residuals have no unit roots. This is an indication that the relevant measured and predicted from labour force time series are likely cointegrated. Such a result for the cumulative curves is not a surprising one – the residual error must have a zero mean despite both series are integrated of order 2.

The Johansen test for cointegration supports the conclusion from the annual residual – the annual measured and predicted curves are cointegrated. The trace statistics gives cointegration rank 1 for two variables. We used the following specifications: trend="none", maxlag=3, but the outcome is the same for other trend specifications and maxlag=7. Formally, the Johansen test cannot be applied to $I(2)$ series and we did not test the cumulative curves.

As mentioned above, small samples usually do not provide statistical estimates and inferences with the desired level of confidence. Fortunately, the OECD also reports quarterly estimates of inflation and labour force. As a rule, monthly and quarterly data are noisy because of measurement errors. For the Canadian time series the overall measurement accuracy is not poor and we have obtained the estimates of coefficients in the linear link between the annual change rate of the GDP deflator for each quarter (annualized Q/Q) and $l_t$:

$$DGDP_t = 3.0 l_{t-8} - 0.0045;\ t \geq 1989$$
$$DGDP_t = 2.0 l_{t-8} - 0.0020;\ t \leq 1989 \qquad (7)$$

where the lag is 8 quarters and the break year is 1989. The change in the lag and break is likely associated with extremely high volatility of quarterly estimates. Thus, this model is a crude one. The change in monetary policy did introduce a tangible disturbance in the link between inflation and labour force. Figure 9 presents the quarterly curves for observed and predicted inflation, both smoothed with MA(8). The resemblance is relatively good.

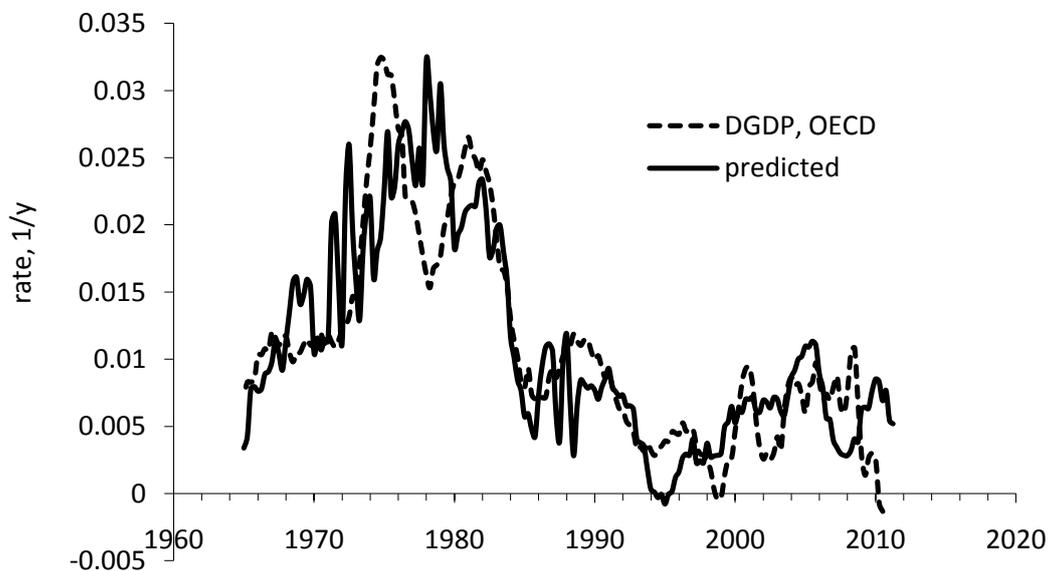

Figure 9. Modelling the quarterly DGDP estimates, both curves are smoothed with MA(8).

We have tested the smoothed time series for cointegration. In the Engle-Granger test for cointegration, the residual error of linear regression should not have unit roots. Figure 10 depicts the model residual, which we consider as an equivalent of the regression residual error.

For 188 readings, the augmented Dickey-Fuller (DF) test $z(t)=-4.37*$ with the 1% critical value of -3.48. The DF-GLS test rejects (for 1% critical value) the null of a unit root for lags 1, 5 and 6 (quarters). The Phillips-Perron test for unit roots gave $z(\rho)=-35.95*$ and $z(t)=-4.50*$, with the 1% critical value of -20.07 and -3.48, respectively. Therefore, all tests for unit roots prove that the predicted time series is cointegrated with the observed one. The Johansen test confirms the presence of a cointegration relation. Econometrically, there exists a long term equilibrium relation between the rate of inflation and the change in labour force in Canada with a break around 1990. This makes the above results of linear regression unbiased.

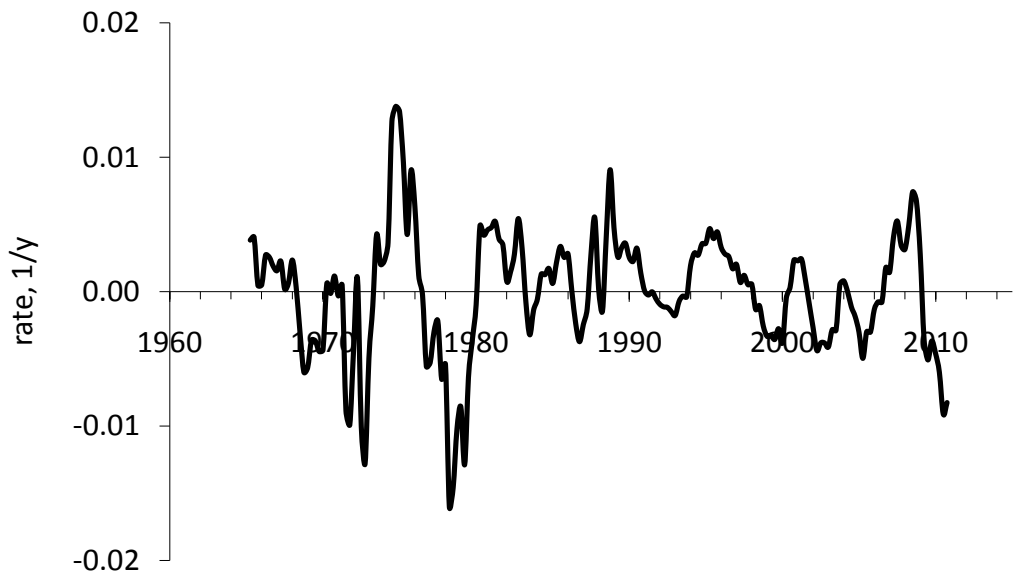

Figure 10. The model residual from Figure 9.

The overall consumer price index allows corroboration of the results obtained for the GDP deflator. We apply the same technique to the annual readings of CPI inflation reported by the OECD. The minimization procedure with the start break year in 1991 (the change in monetary policy) and zero lag (varied between zero and 5 years) has finally determined the data-driven break year in 1991 and the lag of 3 years:

$$CPI_t = 2.682l_{t-3} - 0.0035;\ t<1991$$
$$CPI_t = 0.625l_{t-3} + 0.0104;\ t\geq1991 \qquad (7)$$

Considering the difference between DGDP and CPI in Figure 1 the change in lag is not a surprise. At the same time, the break year fits the introduction of inflation targeting. Figure 11 depicts the annual and cumulative curves; the latter also includes a model without structural break. The annual curves are smoothed with MA(3) and demonstrate a very high degree of similarity for the period between 1964 an 2010 with $R^2=0.89$. For the annual estimates, $R^2=0.68$ and for the cumulative ones $R^2=0.999$. These estimates are not bogus when the annual time series are cointegrated.

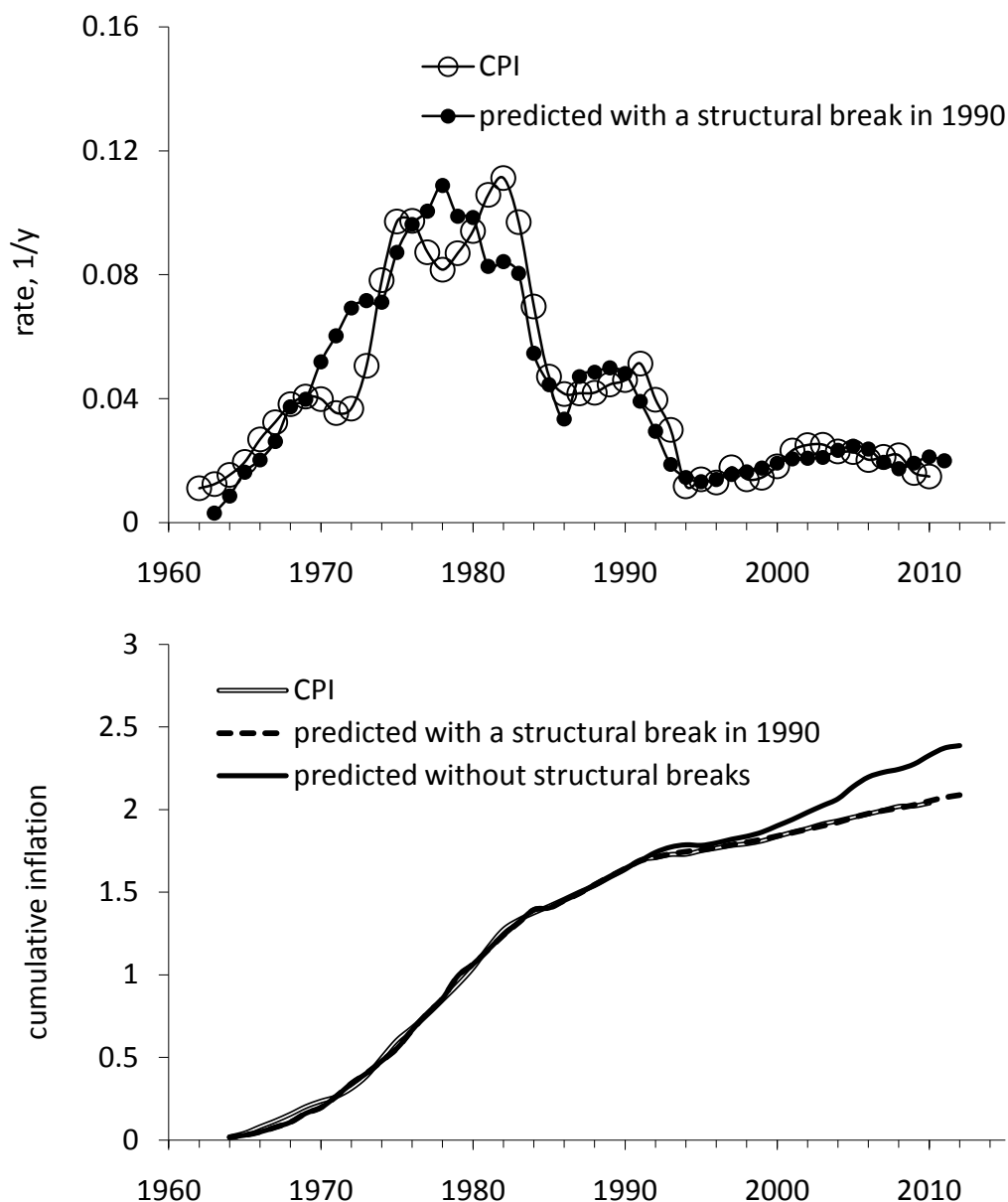

Figure 11. Modelling the annual and cumulative CPI inflation as a function of the change rate of labour force. The structural break was found in 1991. Both annual curves are smoothed with MA(3).

There are 47 readings between 1964 and 2010. We have tested the differences between the dynamic and cumulative curves, i.e. the model residuals, for unit roots. For the annual residuals, the augmented Dickey-Fuller test is -5.83* with the 1% critical value of -3.61, i.e. the null of a unit root is rejected. The Phillips-Perron tests gives $z(\rho)=-31.83*$ (-18.63) and $z(t)=-5.78*$ (-3.61). The DF-GLS test rejects the null for all lags between 1 and 5. Hence, there is no unit root in the model residual, i.e. the rate of CPI inflation and the change in labour force are cointegrated, when one introduces a structural break in the cointegrating relation in 1991. The Johansen test gives rank 1 and confirms the presence of one cointegrating relation between the measured and predicted inflation. The driving force leads by 3 years creating a natural forecast horizon of the same length. The standard deviation in the predicted series (i.e. a three-year-ahead inflation estimate) is $0.024y^{-1}$. This value is smaller than the RMSE of the naïve prediction (Atkeson and Ohanian 2001) at the same 3-year horizon, $0.028y^{-1}$. When one smoothes the predicted inflation series with MA(3) the forecast horizon falls to 2 years. Then the RMSFE of our model is only $0.018y^{-1}$ compared to $0.023y^{-1}$ for the naïve prediction at a 2-year horizon.

Taking into account the cointegrating relations estimated for the DGDP and CPI series one can conclude that the change in labour force has been driving inflation (at least) since the beginning of 1960s. The structural break associated with the introduction of inflation targeting definitely induced shifts in all coefficients, but did not change the linear functional dependence and the lag of inflation. This finding may be interpreted as a shift from one stationary regime of the Canadian economy to another regime, also a stationary one.

All in all, the new monetary policy has affected inflation and unemployment, and both in a desired direction. This confirms the correctness of the Lucas critique. However, both variables are still driven by the change in labour force. Moreover, the joint effect of inflation targeting is zero when the generalized model is applied, i.e. the change in unemployment is fully compensated by the change in inflation a year (or three years) later when equation (3) or (4) is applied.

## 4. The generalized model

In Sections 2 and 3, we have estimated several individual links between labour force, unemployment and inflation. Both individual relations to labour force are cointegrated, as the Engle-Granger and Johansen tests have shown. However, there was a structural break in 1991 induced by the introduction of new monetary policy. In this case, the estimation of a generalized model is a mandatory one. Since inflation lags behind the rate of unemployment and the change in labour force we have estimated model (4) for DGDP and CPI separately:

$$DGDP_t = 3.70l_{t-1} + 0.55u_{t-1} - 0.076$$
$$CPI_t = 3.40l_{t-3} + 0.55u_{t-3} - 0.073 \qquad (8)$$

Coefficients in (8) are similar for both measures of inflation with a little higher influence of labour force on the DGDP. This might be a result of the higher volatility, but the effect of $l_t$ is compensated by a slightly lower intercept of -0.076. The influence of unemployment is essentially the same. The most important finding is that there is no sign of the 1991 structural break in (8) and one equation covers the entire period between 1965 and 2010. This is an obvious consequence of the balance between inflation and unemployment in (8). When the rate of unemployment falls by 3.6% per year the rate of inflation also drops by 0.55*3.6%= 2%. The estimated value of inflation fall after 1991 is 1.9% per year.

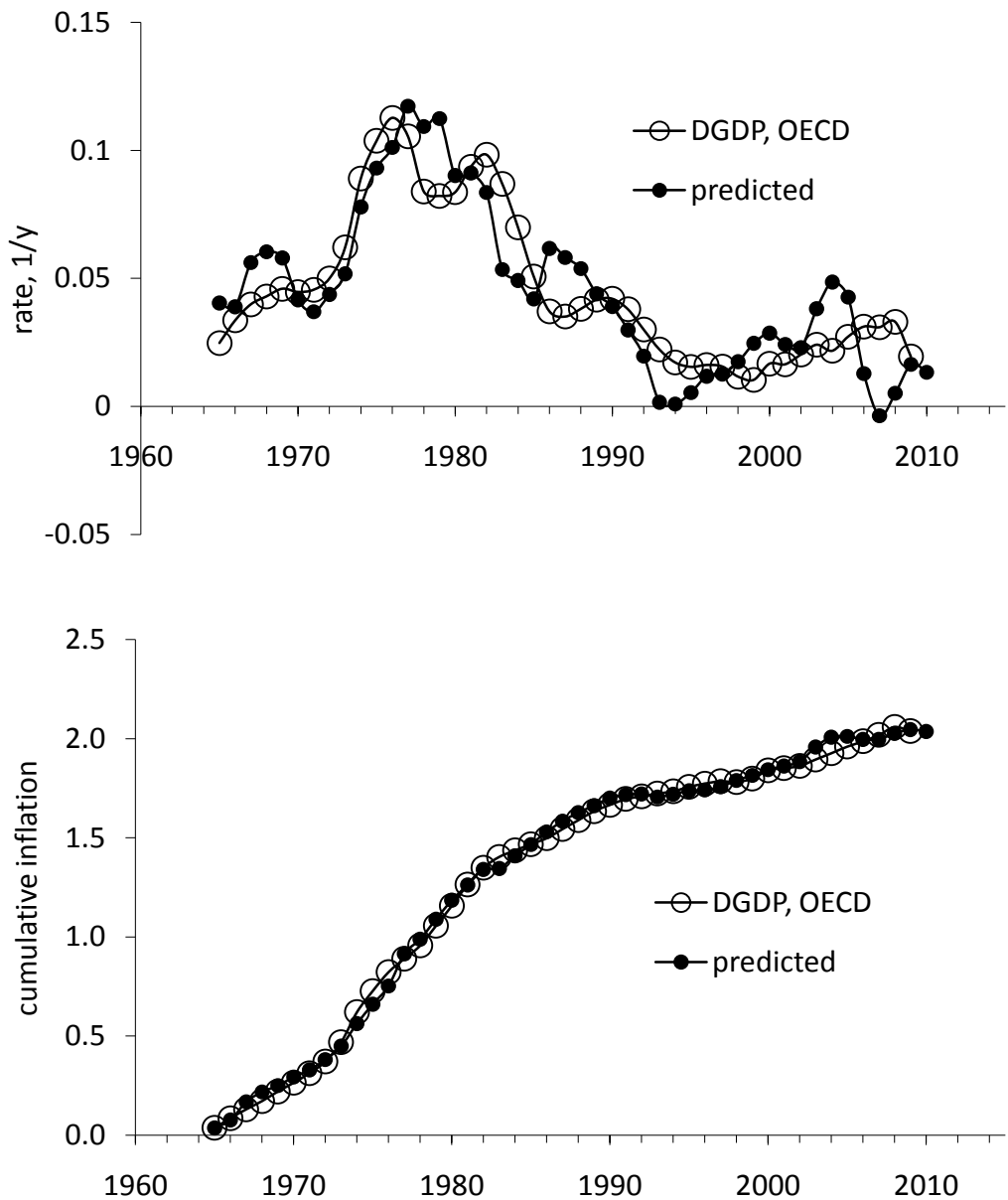

Figure 12. *Upper panel*: Illustration of the generalized relation between inflation, unemployment and the change rate of labour force level in Canada. The GDP deflator is modelled using the change rate of labour force level and unemployment. Both series are smoothed with MA(3). *Lower panel*: These cumulative curves were used to estimate all coefficients in (8).

Figures 12 and 13 present the measured and predicted inflation. The annual series are characterized by $R^2$=0.52 in both cases. In Figure 12, we have smoothed the annual curves with MA(3) and the curves are very close to each other. The difference between the annual series has no unit roots as the augmented Dickey-Fuller (-5.48*) and the Phillips-Perron ($z(\rho)$=-34.47* and $z(t)$=-5.47*) tests show. Thus, there is no unit root in the model residual.

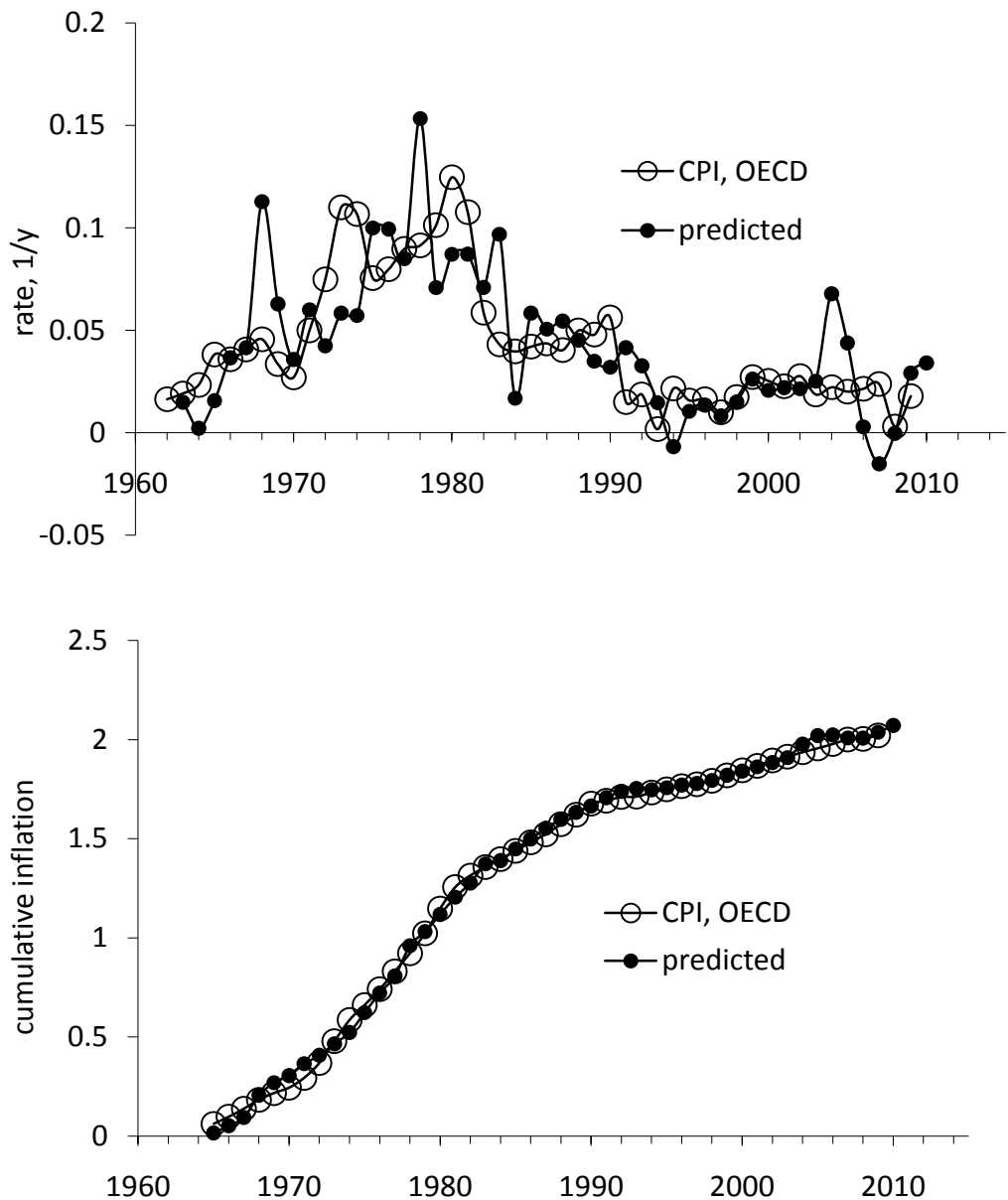

Figure 13. *Upper panel*: Illustration of the generalized relation between inflation, unemployment and the change rate of labour force level in Canada. The CPI inflation is modelled using the change rate of labour force level and unemployment. *Lower panel*: These cumulative curves were used to estimate all coefficients in (8).

In Figure 13, we did not smooth the annual curves in order to demonstrate the reason for a relatively low $R^2$. As discussed in Section 1, there are artificial breaks in the units of labour force and unemployment measurements in 1967 and 1976. One can also suggest that there was a major revision in 1982. After 2005, there is a high-amplitude oscillation potentially catastrophic for any quantitative modelling. All these spikes deteriorate the result of linear regression. However, there is no unit root in the difference between the observed and predicted rate of inflation, as the ADF (-5.47*) and the PP (-35.78* and -5.49*) tests demonstrate.

**Conclusion**

The Lucas critique was correct. The manipulations associated with the introduction of the new monetary policy in 1991 produced a substantial effect on the long-term equilibrium relation between the rate of price inflation and the change in labour force. Amazingly, the monetary policy had a highly positive side effect of a lowered unemployment. In 2010, the rate of unemployment would be around 12% without the inflation targeting. In France, the effect of a similar monetary policy, adapted by the Banque de France in 1995, is opposite – lowered price inflation resulted in the rate of unemployment as high as 12% or ~5 percentage points above the rate without the new policy (Kitov 2007a).

All in all, the rate of price inflation and unemployment in Canada is a one-off function of the change in labour force. This conclusion validates the previous model for Canada and the models for many developed countries: the U.S., Japan, Germany, France, Italy, Canada, the Netherlands, Sweden, Austria, and Switzerland.

Overall, we have established that there exist long term equilibrium relations between the rate of labour force change and the rate of inflation/unemployment. The level of statistical significance of these cointegrating relations in the absence of AR-terms allows us to consider these links as deterministic ones, as adapted in physics. This does not make the rate of unemployment and inflation non-stochastic time series. The change in labour force includes a strong demographic component, and thus, is stochastic to the extent the evolution of population in a given country is stochastic. Since the level of labour force is a measurable value one does not need a data generating process in order to describe its stochastic properties – they are obtained automatically with routine measurements.